\documentclass[preprint,showpacs,preprintnumbers,amsmath,amssymb]{revtex4}

\def\lapprox{\,\raise0.4ex\hbox{$<$}\kern-0.8em\lower0.7ex\hbox{$\sim$}\,}
\def\gapprox{\,\raise0.4ex\hbox{$>$}\kern-0.8em\lower0.7ex\hbox{$\sim$}\,}
\usepackage{bm}

\begin{document}
\bibliographystyle{prsty}

\centerline{\large\bf Conduction-electron spin resonance
in two-dimensional structures}

\vskip 7mm

{ \centerline{Victor M. Edelstein} }

\centerline{\it Institute for Solid State Physics of RAS,
Chernogolovka, 142432 Moscow District, Russia}

\vskip 10mm{}~{}~{}~{}~{}~{}
\parbox{14.2cm} {\rm\small

The influence of the conduction-electron spin magnetization density,
induced in a two-dimensional electron layer by a microwave electromagnetic
field, on the reflection and transmission of the field is considered.
Because of the induced magnetization and electric current, both the electric
and magnetic components of the field should have jumps on the layer.
A way to match the waves on two sides of the layer, valid
when the quasi-two-dimensional electron gas is in the one-mode state,
is proposed. By following this way, the amplitudes of transmitted and
reflected waves as well as the absorption coefficient are evaluated.

\smallskip
\noindent PACS number(s): 73.43.Lp, 75.30.Ds, 76.30.-v, 76.30.Pk}\\
\vskip 10mm
Highlights: \\1. Special matching conditions are needed to evaluate
spin-resonance absorption of 2D conductor;\\
2. Magnetic field acting on electron spins differs from that of incident wave.

\vskip 15mm

\centerline{\bf \uppercase\expandafter{\romannumeral1}.
INTRODUCTION}

\vskip 5mm

Electron spin resonance has long been used to determine {\it g} factors
and the longitudinal and transversal relaxation times $T_{1}$ and
$T_{2}$ providing information about electron band structure and allowing
to investigate interactions responsible for spin-flip transitions \cite{Slicht}.
This method has acquired an enhanced actuality nowadays because a growing
interest in the spin dynamics in two-dimensional (2D) electron systems,
which are potentially important for spintronics applications \cite{Zutic,Wu}.
For a long time it was thought that the direct observation of the
conduction-electron spin resonance (CESR) in 2D structures is impossible because
of small number of current carriers. A breakthrough in this field are recent
works \cite{Wila-1,Wila-2,Tyry,Schul,Wila-3,Gopa,Wolf} where the CESR in
some 2D semiconductor structures was detected by means of the microwave
absorption. The idealness and hence the conductivity of such structures
can be high so that the field acting on electron spins can differ appreciably
from that of incoming wave because of the field of the electric current
excited by the wave. Despite the theory of the spin resonance excitation in
bulk conductors is well elaborated (see, e.g., Refs. [11] and references
therein), an analogous theory for 2D conductors, to the best of the
author knowledge, is still lacking. The purpose of the present note is
to fill in this gap.

\vskip 15mm

\centerline{\bf \uppercase\expandafter{\romannumeral2}.
PROBLEM STATEMENT AND RESULTS}

\vskip 5mm

A feature of this problem
which impedes the immediate application of standard methods, consists
in the following. Let the quasi-2D layer aligned along an $x-y$ plane is
placed at position $z=0$ between two dielectrics with the permittivities
$\epsilon_{1}$ ($z<0$) and $\epsilon_{2}$ ($z>0$), and $z$-axis points
"upward" to the dielectric~2. Within the frame of classical electrodynamics,
properties of a conducting medium enter the Maxwell equations through the
material constitutive relations \cite{LL8}, which in the case under study
have the form
\begin{equation}                                             
{\bf J}=\hat{\sigma}{\bf E}\,, \hspace{2mm} {\bf M}=\hat{\chi}{\bf H},
\end{equation}
\noindent
where $\sigma$ and $\chi$ are tensors of the electric conductivity and the
magnetic susceptibility, respectively. In the following all quantities are
assumed to have the time dependence $e^{-i\omega t}$. The great difference
between the width of the conducting layer $d$ and the wavelength $\lambda
=2\pi/q_{0}$ ($q_{0}=\omega/c$) of microwave radiation urges one to treat
the layer as strictly two-dimensional sheet so that ${\bf J}({\bf r};t)
=\delta(z){\bf J}_{S}(x,y;t)$ and ${\bf M}({\bf r};t)=\delta(z){\bf M}_{S}
(x,y;t)$. Then, from the Amper low
\begin{equation}                                             
\nabla\times {\bf H}_{\omega}=-iq_{0}\epsilon{\bf E}_{\omega}
-\frac{4\pi}{c}{\bf J}_{\omega},\hspace{1mm}
\end{equation}
\noindent it follows the usual expression for the jump of the magnetic
field on the sheet
\begin{equation}                                              
\hat{\bf n}\times ({\bf H}_{\omega,\,2}-{\bf H}_{\omega,\,1})
=\frac{4\pi}{c}{\bf J}_{\omega,\,S}\,
\end{equation}
\noindent where ${\bf n}=\hat{\bf z}$ and ${\bf H}_{\omega,\,1}$ and
${\bf H}_{\omega,\,2}$ are the values of the magnetic field on the lower
and upper sides of the sheet, respectively. Accordingly, from the
Faraday low
\begin{equation}                                             
\nabla\times {\bf E}_{\omega}=iq_{0}\left({\bf H}_{\omega}
+4\pi{\bf M}_{\omega}\right)
\end{equation}
\noindent it follows
\begin{equation}                                             
\hat{\bf n}\times ({\bf E}_{\omega,\,2}-{\bf E}_{\omega,\,1})
=4\pi iq_{0}{\bf M}_{\omega,\,S\parallel}\, ,
\end{equation}
\noindent where ${\bf M}_{\omega,\,S\parallel}$ is the parallel
component of the 2D magnetization density. Thus both the jumps of
the electric and magnetic components of the electromagnetic field
on the sheet should be taken into account. If one tries, as usually,
to utilize Eqs. (3) and (5) as boundary conditions for matching the
fields above and below the sheet, an ambiguity occurs --  the jumps
make undefined the values of ${\bf E}$ and ${\bf H}$ which should be
used in Eqs. (1). Thus, the inequality $d\ll \lambda$ does not allow
one to consider the system as a strictly 2D sheet from the very
beginning. Therefore, we will first consider $d$ as small but finite
quantity, trying to find an additional property of the 2D conductor,
which could lift the ambiguity mentioned, and take the limit
$d/\lambda\to 0$ on a later stage.

This additional property, which the following consideration depends on,
is the assumption that the electron gas is in the one-mode state, i.e.,
all electrons occupy only the ground state in the confinement potential
forming the 2D gas. Such a situation is usual in semiconductor
heterostructures and conducting surfaces and interfaces of oxide insulators.
It will be shown below that at the normal incidence of the wave on the
one-mode gas the 'averaged' fields ${\bf E}_{\omega,\,av}
=\frac{1}{2}({\bf E}_{\omega,\,1}+{\bf E}_{\omega,\,2})$ and
${\bf H}_{\omega,\,av}=\frac{1}{2}({\bf H}_{\omega,\,1}+{\bf H}_{\omega,\,2})$,
where ${\bf E}_{\omega,\,1,2}$ (${\bf H}_{\omega,\,1,2}$) are the limit values
of the electric (magnetic) field on the lower and upper sides of the layer,
respectively, should be substituted into the right-hand sides of Eqs. (3)
and (5). Thus, Eqs. (3) and (5) should take the form

\begin{equation}                                              
\hat{\bf n}\times ({\bf H}_{\omega,\,2}-{\bf H}_{\omega,\,1})
=\frac{4\pi}{c}\hat{\sigma}\left(\frac{{\bf E}_{\omega,\,1}
+{\bf E}_{\omega,\,2}}{2}\right)\, ,
\end{equation}
\begin{equation}                                             
\hat{\bf n}\times ({\bf E}_{\omega,\,2}-{\bf E}_{\omega,\,1})
=4\pi iq_{0}\hat{\chi}\left(\frac{{\bf H}_{\omega,\,1}
+{\bf H}_{\omega,\,2}}{2}\right).
\end{equation}
\noindent The standard method supplemented with this matching conditions
becomes well defined and straightforwardly gives rise to the following
results. The amplitudes of reflection $T_{re}$ and transmission $T_{tr}$
have the form
\begin{eqnarray}                                        
T_{re}=\frac{N_{re}}{D}\, ,T_{tr}=\frac{N_{tr}}{D}\,,
\hspace{4cm} \nonumber   \\
N_{re}=\left(n_{2}-n_{1}+\frac{4\pi\sigma_{\omega}}{c}\right)
+2\pi iq_{0}\chi_{\omega}\left[2n_{1}n_{2}+\frac{2\pi\sigma_{\omega}}{c}
(n_{1}-n_{2})\right]\,,\nonumber  \\
N_{tr}=2n_{2}+2\pi iq_{0}\chi_{\omega}\frac{4\pi\sigma_{\omega}}{c}n_{2}
 \hspace{3cm} \\
D=\left(n_{2}+n_{1}+\frac{4\pi\sigma_{\omega}}{c}\right)-2\pi iq_{0}
\chi_{\omega}\left[2n_{1}n_{2}
+\frac{2\pi\sigma_{\omega}}{c}(n_{1}+n_{2})\right]\,,  \nonumber
\end{eqnarray}
\noindent while the absorption coefficient, with the accuracy up to terms
linear in $\chi$, is
\begin{eqnarray}
A=\frac{N}{Z}, \hspace{4cm} \nonumber   \\
N=\frac{4\pi\sigma'_{\omega}}{c}n_{1}+4\pi q_{0}n_{1}n_{2}^{2}
\chi''_{\omega}\left[1+\frac{16\pi\sigma'_{\omega}}{cn_{2}}
+\left(\frac{2\pi\sigma'_{\omega}}{cn_{2}}\right)^{2}+
\left(\frac{2\pi\sigma''_{\omega}}{cn_{2}}\right)^{2}\right],\\    
Z=\left(n_{1}+n_{2}+\frac{2\pi\sigma'_{\omega}}{c}\right)^{2}+
\left(\frac{2\pi\sigma''_{\omega}}{c}\right)^{2}.
\hspace{2cm}      \nonumber
\end{eqnarray}
\noindent These equations have been written for the fields with
the circular polarization ${\bf e}_{+}=\frac{1}{\sqrt{2}}
({\bf e}_{x}+i{\bf e}_{y})$ when ${\bf E}=E_{(-)}{\bf e}_{+}$,
${\bf H}=H_{(-)}{\bf e}_{+}$, ${\bf M}=M_{(-)}{\bf e}_{+}$,
${\bf J}=J_{(-)}{\bf e}_{+}$ and the constitutive relations
have the form $M_{(-)}=\chi_{\omega}^{(+)}H_{(-)\,av}$
with $\chi_{\omega}^{(+)}=\chi_{xx}(\omega)+i\chi_{xy}(\omega)$
and $J_{(-)}=\sigma_{\omega}^{(+)}E_{(-)\,av}$ with
$\sigma_{\omega}^{(+)}=\sigma_{xx}(\omega)+i\sigma_{xy}(\omega)$.
Also the following notations have been used: $n_{1,2}=\sqrt{\epsilon_{1,2}}$
is the refraction index,
$\chi''_{\omega}=\Im \chi_{\omega}^{(+)}$, $\sigma_{\omega}
=\sigma_{\omega}^{(+)}$, $\sigma'_{\omega}=\Re \sigma_{\omega}^{(+)}$,
and $\sigma''_{\omega}=\Im \sigma_{\omega}^{(+)}$.
Near the frequency $\omega_{res}$ of the CESR one gets \cite{Edel}
$\chi^{(+)}_{\omega}\cong\chi_{0}\frac{\pi}{m}N(\epsilon_{F})
\frac{-\omega}{\omega-\omega_{res}+\frac{i}{T_{2}}}$,
 where $\chi_{0}=\frac{m}{\pi}
\left(\frac{g\mu_{B}}{2}\right)^{2}$ is the static susceptibility
of 2D degenerate electron gas and $N(\epsilon_{F})$ is the density of
states for a single spin. Eqs. (8) and (9)
show that at $\sigma/c\ge 1$ the effect of the electric current,
induced by microwave field, on effective magnetic field acting on
electron spins can be appreciable. The derivation of Eqs. (8) and (9)
is quite standard and therefore is not given here. The remaining part
of the paper presents the proof of the above matching conditions.

\vskip 15mm

\centerline{\bf \uppercase\expandafter{\romannumeral3}.
MATCHING CONDITIONS}

\vskip 5mm

So we consider the electron gas which occupies the layer
$-\frac{d}{2}\le z \le \frac{d}{2}$. Two facts follow from the
assumption about the one-mode state of the gas (see Appendix).
The first is that the coordinate dependence of the 3D density of
the current and the magnetization has the factorized form
\begin{equation}                                        
{\bf J}({\bf r},t)=\rho (z){\bf J}_{S}({\bf r}_{\parallel},t),\,
{\bf M}({\bf r},t)=\rho (z){\bf M}_{S}({\bf r}_{\|},t),
\end{equation}
\noindent where ${\bf r}=(x,y,z)=({\bf r}_{\|}\,,z)$, $\rho (z)
=|\psi_{0}(z)|^{2}$, $\psi_{0}(z)$ is the wave-function of the ground
state, and ${\bf J}_{S}$ and ${\bf M}_{S}$ are the 2D densities. At
the normal incidence of the radiation, ${\bf J}_{S}$ and ${\bf M}_{S}$
loose their coordinate dependence. Second fact is that the constitutive
relations (1) take the form
\begin{equation}                                        
J^{i}_{S}(\omega)=\sigma_{\omega}^{ij}\int_{z}\rho(z)
E^{j}(z,\omega),\,
M^{i}_{S}(\omega)=\chi_{\omega}^{ij}\int_{z}\rho(z)
H^{j}(z,\omega)\,,
\end{equation}
\noindent where $\int_{z}=\int dz$.

Consider first the question about the value of the electric field which should
be used in the Ohm's law in the limit $d/\lambda\to 0$. As it is known
[and also seen from Eq. (5)], the major reason for a finite difference
between the electric field on the upper and lower surfaces of the layer
is the magnetization. To make the following explanation more clear the
effect of the external magnetic field is omitted for a while. Consider
a strictly 2D sheet, uniformly filled with the spin magnetization
$\rho(\zeta){\bf m}^{-i\omega t}$, ${\bf m}\perp {\bf e}_{z}$ [${\bf m}$
does not depend on ${\bf r}_{\|}$ at the the normal incidence], which
lies inside the layer at $z=\zeta$, $|\zeta|\le\frac{d}{2}$. By utilizing
the fact that the vector-potential created at the point ${\bf r}$
by the magnetic dipole ${\bm \mu}({\bf r}_{0})$ placed at the point
${\bf r}_{0}$ is given by ${\bf A}({\bf r})=\frac{\mu({\bf r}_{0})
\times({\bf r}-{\bf r}_{0})}{|{\bf r}-{\bf r}_{0}|^{3}}$ \cite{LL2}, one can
show that the vector-potential created by the magnetization of the sheet is
\begin{equation}                                        
{\bf A}({\bf r},t)=e^{-i\omega t}2\pi\rho(\zeta)\left({\bf m}
\times{\bf e}_{z}\right)sign(z-\zeta),
\end{equation}
\noindent so that the vector-potential created by the total magnetization
of the electron layer is given by
\begin{equation}                                        
{\bf A}_{\omega}(z)=2\pi\left({\bf m}\times{\bf e}_{z}\right)\left[
\int_{-d/2}^{z}\rho(\zeta)d\zeta
-\int_{z}^{d/2}\rho(\zeta)d\zeta\right].
\end{equation}
\noindent The corresponding part of the electric field, ${\bf E}_{\omega}
=\frac{i\omega}{c}{\bf A}_{\omega}$, has the same space dependence.
According to Eq. (11), the electric current induced by {\it this} part of
the field is defined by the expression
\begin{equation}                                        
\int^{d/2}_{-d/2}\rho(z){\bf E}_{\omega}(z)dz\sim
\left({\bf m}\times{\bf e}_{z}\right)
\int^{d/2}_{-d/2}\rho(z)\left[
\int_{-d/2}^{z}\rho(\zeta)d\zeta-\int_{z}^{d/2}\rho(\zeta)d\zeta\right]dz,
\end{equation}
\noindent which equals zero at any function $\rho(z)$. To see this fact
one should consider the second term in Eq.(14) as the double integral
$\int\int dz\,d\zeta\rho(z)\rho(\zeta)$ over the triangle region
$-\frac{d}{2}\le z,\zeta\le \frac{d}{2},\, z\le \zeta$, perform the
$z$-integration first, and then change variables $z\leftrightarrow \zeta$.
Thus, in the case of one-mode conductor, when the space dependence of
both the current and the magnetization densities are defined by the same
function $\rho(z)$, that part of the electric field, which is induced by
the oscillating magnetization, does not give rise to the electric current.
This result is the central point in the derivation of Eqs. (6,7).
Within the layer, the total electric field ${\bf E}_{\omega}(z)$ can be
represented as the sum of a slow-space-varying component ${\bf E}_{sl}(z)$,
whose scale of variance is the wave length $\lambda$, and the
fast-space-varying component ${\bf E}_{f}(z)$ due to the magnetization
[which yields the finite jump on the layer in the limit $d/\lambda\to 0$],
as ${\bf E}_{\omega}(z)={\bf E}_{sl}(z)+{\bf E}_{f}(z)$. The fast component
does not participate in the Ohm's law while the slow component is almost
constant within the layer and with the accuracy up to corrections of the
order of $d/\lambda$ can be taken at any point inside the layer, say at
$z_{0}$, $|z_{0}|<\frac{d}{2}$. But, with the same accuracy, we have
${\bf E}_{sl}(z_{0})=\frac{1}{2}\left[\,{\bf E}_{sl}\left(\frac{d}{2}\right)
+{\bf E}_{sl}\left(\frac{-d}{2}\right)\right]$. Then, since $\int_{z}\rho(z)=1$,
we came to the Ohm's law with the anzatz ${\bf E}_{\omega}={\bf E}_{sl,\,av}$.
But because for the fast component one has
${\bf E}_{f}\left(\frac{d}{2}\right)+{\bf E}_{f}\left(\frac{-d}{2}\right)=0$,
one may change the 'average' of the {\it slow} field ${\bf E}_{sl,\,av}$ by
the 'average' of the {\it total} field ${\bf E}_{\omega,\,av}$. After then
one can take the limit $d/\lambda\to 0$ thereby obtaining Eq. (6).

The Eq. (7) can be proved
following the same lines: by using the fact that the magnetic field created at the
point ${\bf r}$ by the current with the density ${\bm j}({\bf r}_{0})$ is given by
${\bf H}({\bf r})=\frac{1}{c}\int_{{\bf r}_{0}}\frac{{\bf j}({\bf r}_{0})\times
({\bf r}-{\bf r}_{0})}{|{\bf r}-{\bf r}_{0}|^{3}}$ \cite{LL2}, one can find explicitly
the magnetic field inside the layer induced by the electric current and to show that
this magnetic field does not give rise to a contribution to the magnetization.
Since at the presence of an applied constant magnetic field the form of Maxwell's
equations for ${\bf e}_{+}$ and ${\bf e}_{-}$ circular polarizations coincides with
that at the absence of the field, the proof presented holds in those cases as well.

\vskip 10mm

\centerline{\bf \uppercase\expandafter{\romannumeral4}.
CONCLUSIONS}

\vskip 5mm

In summary, it has been considered one problem of the macroscopic
electrodynamics of 2D paramagnetic conductors, in which it is 
necessary to take care of jumps of both the electric and magnetic 
component of the electromagnetic field on the conductor. Namely, 
the reflection of electromagnetic wave with a frequency near the 
CESR. Some way to solve this problem has been pointed 
out. Physical sense of the matching
conditions found, Eqs. (6) and (7), is that if the electron gas
is in the one-mode state one can disregard the jump of the electric
field by evaluating the jump of the magnetic field and, quite
analogously, one can disregard the jump of the magnetic field by
evaluating the jump of the electric field. The results obtained,
Eqs. (8) and (9), describe
the contribution of the CESR, as well as the cyclotron resonance,
to the transmission/reflection amplitudes and to the absorption
coefficient. In the limit of the very small conductivity, Eqs. (8,9)
describe the transmission through a paramagnetic insulator, while at
$\chi_{\omega}\to 0$ we recover the absorption only due to the
cyclotron resonance \cite{Chiu}. Note that the statement of the
problem of the CESR excitation adopted in this paper holds for 2D
structures with small up-down asymmetry, like Si/SiGe quantum wells
investigated in works \cite{Wila-1,Wila-2,Tyry,Wila-3,Gopa,Wolf}. In
such structures the ESR reveals itself in the ordinary fashion -
through $\chi_{\omega}$. In semiconductor structures with strong
Rashba spin-orbit coupling, the CESR can reveal itself more
pronouncedly through $\sigma_{\omega}$ than through $\chi_{\omega}$.
This is the case, e.g., in AlAs quantum wells, as it has been shown
experimentally \cite{Schul} and theoretically \cite{Edel}.
In those cases, the spin susceptibility (and hence the jump of the
electric field on the layer) plays a minor role and can be disregarded.

Note that special matching conditions discussed above are not peculiar
to electrodynamics of conventional 2D systems. They are also required
for an evaluation of the paramagnetic response of 2D conductors with the
Weil-type Hamiltonian.

\vskip 3mm
This work was supported by grant 14-12-01290 of Russian Science Foundation.

\vskip 3mm

\appendix
\section{}

\vskip 5mm

In this Appendix it is shown how one can derive Eqs. (10) and (11). A straightforward
way is to use the Kubo formalism \cite{AGD} which yields, for
example, for the 3D current density
\begin{equation}                                        
J_{i}({\bf r}|\omega)\sim\int_{\bf r'}\Pi_{ij}^{R}({\bf r},{\bf r}'|\omega)
E_{j}({\bf r}'|\omega),
\end{equation}
\noindent where $\Pi_{ij}^{R}({\bf r},{\bf r}'|\omega)$ is the retarded velocity
correlation function, which in the Feynman diagram language is given by a sum of
loop ladder diagrams. For the conducting layer in the one-mode state and at the
normal incidence of the radiation, when the electric field is parallel to the
layer and is a function of only $z$,  the radiation cannot induce transitions
into excited states in the confinement potential. Under these conditions, the 3D
one-electron Green's function $G^{3D}$ relevant to the problem can be expressed
through the 2D Green's function $G_{S}$ as
\begin{equation}                                        
G^{3D}({\bf r},{\bf r'}|\epsilon)=\psi_{0}(z)
G_{S}({\bf r}_{\|},{\bf r'}_{\|}|\epsilon)\psi_{0}(z').
\end{equation}
\noindent Consider the contribution to $\Pi_{ij}^{R}$ of the simplest diagram (which
is the loop without impurity insertions). Since the parallel components of the velocity
operator $\hat{\bf v}({\bf r}_{\|})$ do not act on the wave functions of
perpendicular motion $\psi_{0}(z)$ and $\psi_{0}(z')$, this diagram yields
\begin{equation}                                        
T\sum_{\epsilon}|\psi_{0}(z)|^{2}Tr\left\{\hat{v}_{i}({\bf r}_{\|})
G^{R}_{S}({\bf r}_{\|},{\bf r'}_{\|}|\epsilon+\omega)
\hat{v}_{j}({\bf r'}_{\|})G^{A}_{S}({\bf r'}_{\|},{\bf r}_{\|}|\epsilon)\right\}
|\psi_{0}(z')|^{2}
\end{equation}
\noindent so that the corresponding contribution to the current $J_{j}({\bf r}|\omega)$
is proportional to
\begin{equation}                                        
|\psi_{0}(z)|^{2}\int_{{\bf r'}_{\|},z'}T\sum_{\epsilon}Tr\left\{\hat{v}_{i}({\bf r}_{\|})
G^{R}_{S}({\bf r}_{\|},{\bf r'}_{\|}|\epsilon+\omega)
\hat{v}_{j}({\bf r'}_{\|})G^{A}_{S}({\bf r'}_{\|},{\bf r}_{\|}|\epsilon)\right\}
|\psi_{0}(z')|^{2}E_{j}(z'|\omega)
\end{equation}
\noindent It is seen that this expression reproduces the form of Eqs. (10) and (11).
It can be straightforwardly checked that the same is true with respect to
contributions of all other diagrams. The validity of the 'magnetic' parts of
Eqs. (10) and (11) can be proved quite analogously.

\begin{equation}
{\bf J}({\bf r},t)=e{\bf v}(t)|\psi({\bf r}-{\bf R}(t))|^{2}
\end{equation}
\begin{equation}
{\bf F}=\int e|\psi({\bf r})|^{2}{\bf E}({\bf r})d^{3}{\bf r}
\end{equation}
\begin{equation}
{\bf J}_{S}(\omega)=\sigma_{\omega}\int_{z}\rho(z)
{\bf E}(z,\omega),\,
{\bf M}_{S}(\omega)=\chi_{\omega}\int_{z}\rho(z)
{\bf H}(z,\omega)\,,
\end{equation}

\begin{figure}[h]

\end{figure}

\end{document}